# Multi-agents Architecture for Semantic Retrieving Video in Distributed Environment

Yasser El Madani El Alami
Sidi Mohammed Ben Abdellah University
Faculty of Science
Fez, Morocco
yasser.alami@hotmail.fr

El HabibNfaoui
Sidi Mohammed Ben Abdellah University
Faculty of Science
Fez, Morocco
elhabib.nfaoui@usmba.ac.ma

Omar El Beqqali
Sidi Mohammed Ben Abdellah University
Faculty of Science
Fez, Morocco
omar.elbeqqali@usmba.ac.ma

*Abstract*—This paper presents an integrated multi-agents architecture for indexing and retrieving video information. The focus of our work is to elaborate an extensible approach that gathers –a priori- almost of the mandatory tools which palliate to the major intertwining problems raised in the whole process of the video lifecycle (classification, indexing and retrieval). In fact, effective and optimal retrieval video information needs a collaborative approach based on multimodal aspects. Clearly, it must to take into account the distributed aspect of the data sources, the adaptation of the contents, semantic annotation, personalized request and active feedback which constitute the backbone of a vigorous system which improve its performances in a smart way.

## I. INTRODUCTION

Since the emergence of the Internet,the rapid development realizedin computing systems sector (pc, smartphones, tablets, etc.) and the progress in mass storage technologies, the amount of information turned digitalhas witnessed an exponential grow (figure1). In fact, over 92% of new information generated is stored digitally [1] with nearly 1 zettabyte information added by year [2] where multimedia information takes the major part. According to the statistics made by YouTube [3] 60 hours of video are uploaded every minute. In addition, in 2016, global IP traffic will reach 1.3 zettabytes per year, the video trafficwill most 55% of total traffic [4].

In addition, it's important to note that, nowadays, multimedia especially audio-visual information takes first podium of the most information that people are looking for. Thus, many media platforms have been made to satisfy people desire and to manage the exponential growing video collection. However, subjective annotations or low level features do not allow users to access the relevant information easily. It's always a hard and tedious task for them. This dephasing between on one hand the limited description of low level features analysis and on the other hand the easiness and the affluence of human interpretation are called "the semantic gap". Referred to Smeulders& al. [5] "the semantic gap is the lack of coincidence between the information that one can extract from the" multimedia material" and the interpretation that the same data have for a user in a given situation".

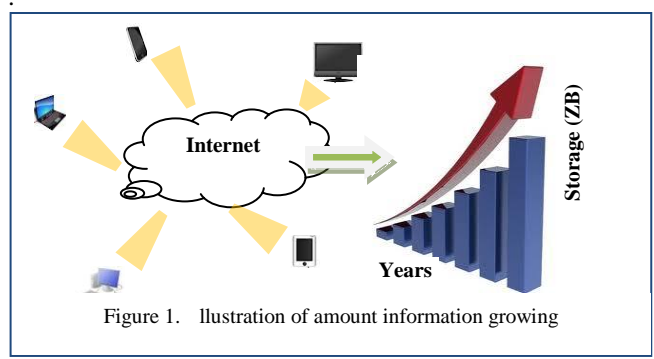

Figure 1. llustration of amount information growing

Matt Roach [21] redefined it as "the lack of coincidence between the formative and cognitive information" where formative information is the shape, form or pattern held within the sequence of multidimensional matrices that make up the video and cognitive information is the information pertaining to 'knowing'. This involves the interpretation of the formative information by a human viewer.

The focus of our work is to propose a reliable architecture that aim to fulfill the semantic gap. It relies not only in video analysis but it's based on collaborative approach which consider user as a primary actor in the retrieval process.

This paper is organized as follow: in section 2 we give a sum-up of related works that have been done in video retrieving; at section 3 we summarize the important works which have been done in system retrieval based on multi-agents system. Then, in section 4, we present our proposed solution. Finally we end this paper with conclusions and our perspectives.

## II. RELATED WORKS

A lot of retrieval information video systems were developed to satisfy the users' needs. Actually, Internet abound in these new tools what takes a large part of users' time navigation. So far, almost of them are based on text. It's the natural and easy approach to build and use. Thenewest approach is based on content. The aim is to fulfill the semantic gap. In the following, we give a sum up of the most important approaches which drive all researches made in this area: Text-based retrieval, Content based retrieval, concept based retrieval and Hybrid approaches.

*A. Text-based video retrieval*

   *1) Manual annotation*

   First, Text-based retrieval is the most popular used paradigm in this domain. Started since 1970s, it was the first way used to index and retrieve document by representing each one by a list of words or keywords which describe its content. The keyword is the atomic unit manually annotated by user such as surrounding text, social tags, closed captions or speech recognition. In general, it's stored, e.g., alphabetically, in the 'index file' [6] or in DBMS to perform retrieval action. Several academic prototypes, such as Medusa [7], Informedia classic [8], and Olive [9], and other online video search engines such as YouTube, Baidu, Blinkx, and Truveo, provide access to video based on text [10].

   *2) MetaData*

   This approach consists of using a model of information that describes any video document. Indeed, metadata includes video characteristics such as video title, date, actor(s), producer(s), video genre, running time and file size, video format, reviews by users and user rating, copyright and ownership information, and so on [11]. User can build a simple query with each meta-data fields or combine it for more accuracy. All most of systems adopting this approach uses "key frame" where they allow users to retrieve video document by request using each of metadata fields. Then, they give an access to key frame or storyboard previews of each video within the whole selected videos which enable users to preview the content itself in a visual way [6]. "Open Video Project" [12] is an example pioneer video retrieval system based on metadata (sponsored by and developed at the Interaction Design Laboratory at the School of Information and Library Science, University of North Carolina Chapel Hill).

   *3) Video text transcription*

   In most of cases, the text included in video gives more relevant information and efficiency that users need. It can be obtained from video with to different manner: speech recognition or character recognition. Text-based video retrieval based on speech transcripts [4] usually used for broadcast news, interviews, political speeches and documentary movies (e.g in English) and so on.It can be obtained from automatic speech recognition (ASR) or from captions available in DVD or certain broadcast TV programs [13]. Olive [9] is a good example of speech recognition which automatically produces indexes from a transcription of the sound track of a program. In contrast, character recognition can be extracted from video by an optical character recognition (OCR) process. Applied to frames' video, it detects word from banners and captions [8].Fishlar News [14] is a video retrieval system which uses video text transcription. The text founded help to segment an individual broadcast into a set of news stories parts and to link them according to their topics.

   The advantage of text-based video retrieval that it's easy to implement because it doesn't require an advanced video analysis and it is satisfied with tools that have reached a maturity level such as request model (SQL request) of DBMS. Unfortunately, this approach isn't being an efficient way to extract relevant video documents due their queries limit to express the real user's needs resulting from their annotations manner. In fact, text based retrieval is usually inaccurate due subjective human perception [15].

   In addition, this method marginalizes content information -features- which are the clue of relevant information access. Indeed, video document contains several features abound in relevant information not well exploited.Moreover, Video retrieval technique based on video text transcription results meets several performance problems on text recognition process: (a) is only applicable in domains with many text inserts, such as news, (b) when the videos originate from non-English speaking countries, such as China [8].

*B. Content-based retrieval*

   Content based video retrieval is an important area of research which inspired several systems. It has been proposed in order to overcome drawbacks of text based retrieval. CBVR consist of the video analysis based on their visual contents (pixels). As known, video is a multimedia sequences comprised of both sound and a series of images [21]. Thus, the image is in the heart of any study conducted on video and the development image research affects systematically progress in indexing and extraction video field. In [16], **Eakins** mentioned three levels of queries in CBIR: low level consists of Retrieving by primitive features such as color, texture, shape or the spatial location of image elements. Intermediate level based on retrieving objects of given type identified by derived features, with some degree of logical inference. High level where retrieving process focus on abstract attributes, involving a significant amount of high-level reasoning about the purpose of the objects or scenes depicted. This includes retrieval of named events, of pictures with emotional or religious significance, etc.

   The process of content-based video retrieval consists of three main tasks (figure): parsing, description and indexing where the first important one is parsing. It consists of dividing the video into individual shots and scenes. A shot is a semantic unit that represents a series of consecutive frames taken by a camera running in continuous time. A Scene is a group of consecutive shots filmed in the same location. Whose boundaries are determined by editing point or where the camera switches on or off [17].This phase can be performed in automatic way by using shot/scene boundary detection algorithms for more details [18], [19]. It consists of the recognition of considerable discontinuities in the visual-content flow of a video sequence [20].

   *1) Summarization*

Referred to [21] Video summarizing try to capture semantic content of a video and present a general highlight of the video in a shorter period of time. It attempts to build a storyboard by reducing time watching while simultaneously keeping meaningful visual content. A storyboard is generated by grouping video keyframes' in the order of appearance in video. Several researches are focused in summarization techniques [22,23,24,25,26,27]. Therefore, it will reduce time wasted when a user is looking for a given video. According to [28] A few seconds of lost keyframes or non-relevant keyframes might not be critical for a user browsing a large number of keyframes within a video. Many systems are based on video summarizing techniques as Open Video Project [12] and VSUM [29]. For example, open Video project allow users to access previewing video document via a set of key frames for a whole video, viewing the video at fast forward speed, or an automatically generated 7-second summary.

*2) Spatio-temporal analysis*

Almost all of works reported in the literature can be considered as an extension of well-known content based image retrieval. In fact, most of CBVR rely on image content understanding and pattern recognition approaches. However previous CBVR researches ignore the temporal aspect of video and not use either the motion of objects or camera motion as part of video analysis, indexing or retrieval. This point involved several researches in spatio-temporal aspect within video [30]. According to [27], SPA attempts to provide on one hand a representation for video components such as frame, shot and scene at different level of abstraction. On the other hand, provide description of the spatial composition among video objects in each frame including directional and topological relations, and temporal composition among frames within shot and sequences. Object motion has been an important feature in spatio-temporal analysis for activity representation in video applications [31][32]. An integrated system for spatio-temporal video retrieval is LucentVision [33] and SEMCOG system [34].

*C. CONCEPT BASED VIDEO RETRIEVAL*

Concept based video retrieval CBVR can be considered as the latest trend in the video retrieval researches. According to [35] It became the most popular approach to bridge the semantic gap. The fundamental idea of CBVR is to define a concept lexicon from human perspective, build a detector based on learning method (supervised learning a priori), and then automatically index the video with the detected concepts [36]. Thus, the user request is mapped to a set of concept and then the relevant videos to these concepts are returned. In brief linguistic cues are used to represent, index and, thus, retrieve the non-linguistic audiovisual content. [37] Estimate that 5000 concepts will be sufficient for an acceptable result while [10] declared 17000 concepts will be better like estimation for an accurate retrieval. [10] Mentioned that several researches drived in this area.

In [38], Smeaton states: "This appears to be the road map for future work in this area." Despite the potential of concept-based video retrieval, however, automatic methods have not yet reached a good performance level. Also, they are not sufficient to solve all retrieving problems. Thus, eventually user involvement is essential.

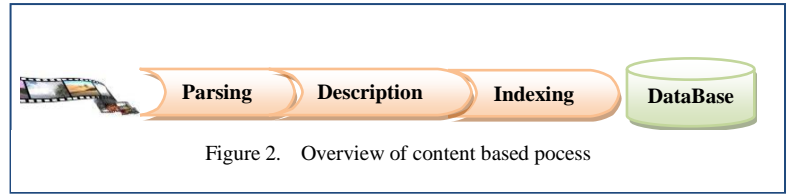

Figure 2.　Overview of content based pocess

*D. Multimedia Ontology*

In order to deal with semantic gap, explicit semantics represented by ontologies has been intensively used in the multimedia retrieval in the last decade [57]. Indeed, Multimedia ontologies were elaborated to provide a high level representation of visual knowledge [58] and at the same time to allow automatic processing over the represented knowledge[59]. In addition, ontologies have the potential to improve the interoperability of different applications producing and consumingmultimedia annotations [60]. The first works that have been done in multimedia ontologies were focused on converting MPEG7 standard to ontology-alike formats[61]such as MPEG-7 Upper MDS, MPEG-7 Tsinaraki[62], and MPEG-7 Rhizomik.More detailscan be found in [63]. Using Mpeg7 is due to the fact that Mpeg7 is an important standard [64] in the multimedia domain for describing content data using low level descriptors .Another example is COMM Ontology [65] which is one of the first references in that direction developed as solution for high level quality multimedia ontology that satisfy a set of requirements such as interoperability and MPEG-7 compliance. The ontology M3O [66] is a second ontology, developed with the aim to provide a pattern that allows accomplishing exactly the assignment of arbitrary metadata to arbitrary media [60]. Figure11 in [60] give a sum up of several multimedia ontologies in the last decade.

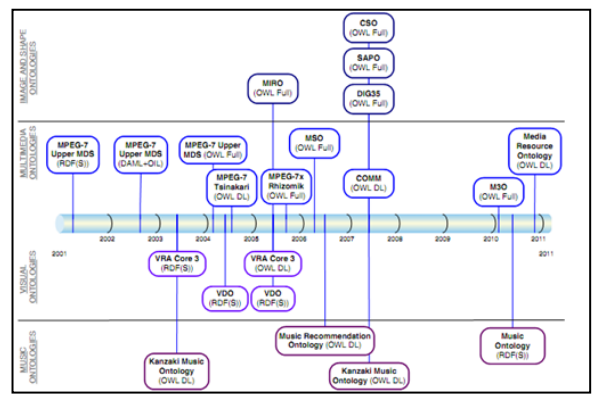

Figure 3.　Time line for the ontologies in the multimedia domain from 2001 to 2011

*E. Hybrid approaches*

Despite the great strides that these approaches have been done, using a single approach does not provide the expected relevance. In fact, each approach deals with a singular problem whereas video is composed from a set of intra-features and extra-features which can have semantic representation for users. Thus, combining approaches will be more accurate and efficient to increase the relevance of results. [39] and [40] are an examples of latter case. They use for example concept-based retrieval to initiate a search, and content-based retrieval similar first result. Other example combines metadata and video summarization such as Físchlár News [28].The following table recapitulates the major video indexing and retrieving approaches.

| Approach | | Principle | Advantages | limitations |
|---|---|---|---|---|
| **Text based retrieval** | Manual annotation | Represent video by a list of words (Subjective annotation) which describe its content. | easy to implement with a mature tools such as RDBMS | Several expressive problems are meeting such as ambiguity. |
| | Metadata | using a model of information that describes any video document | Video description is more objective and requires little analysis of the video content | Limited in terms of supporting a user's information seeking and searching requirements. |
| | Video text transcription | Extract text from video with speech recognition process or optical character recognition process | Rely on advantages of text based retrieval. | Only applicable in domains with many text inserts, such as news |
| **Content based retrieval** | Video summarizing | Generate a semantic storyboard which present a general highlight of the video in a shorter period of time | Reduce time wasted when a user is looking for a given video (browsing time). | It cannot be used alone for an accurate retrieve because user must operate video discrimination himself. |
| | Low level analysis | Video is broken into manageable components such as shot, frame, color or texture. | Approximate content  Extract real features content (shape, color, object...) | Too far from the human cognition. This is not an intuitive way to retrieve |
| | Spatio-temporal analysis | Provide a representation for video components (objects in each frame) throughout video time. | IT's very useful in video surveillance in general in narrow domain. | Have not yet reached a good performance level with full automatic way. More uncertainty in wide domain |
| **Concept based retrieval** | Concept based retrieval | Use automatic lexicon detector in the video. Each lexicon is represented by an visual concept | Maps intuitive human request with visual concept | It hard to descript any human concept. Lexicon detector performance |
| **Hybrid approach based retrieval** | Concept based and content based | Used concept-based retrieval to initiate a search, and content-based to retrieve similar videos | A promising approach to reduce the semantic gap problem | Need a large training set for more effectiveness |
| | Video summarization and metadata | Used metadata for discriminating result set and summarization for quickest navigation in each returned video. | More efficient and less navigation time wasted compared to the full text-based retrieval. | still limited because it does not take into account the content within the video. |

III. MAS FOR RETRIEVING INFORMATION

Since its introduction by distributed artificial intelligence field, the concept of multi-agent systems became a base for the resolving complicated problems having a distributed nature and required a real-time decision without human intervention. Several authors have proposed variant definitions of concept taking inspiration on J.Ferber works such as [41] gathered in his book [42]. For example Jarras et al [43] defines a multi-agent system as distributed system consisting of a set of agents while E.Oliveria et al [44] defines it as a collection of software entities, possibly heterogeneous with their own problem-solving abilities that are able to interact in order to achieve a global goal. It can also be defined as a society of agents apprentices, autonomous, having a limited understanding of the global environment, with a set of resources and acting on behalf of a third party in order to achieve the objective they have been assigned.

The fundamental and atomic part within the system is the "agent". It's defined in [45, 46] as a computer system, situated in an environment, and acts of an autonomous and flexible to achieve the objectives for which it was designed. More detailis present onJ.Ferber's book [42].

MAS have been successfully applied to a number of complex problems and have proven their performance such as information retrieval domain. As defined in [47], information retrieval is finding document of an unstructured nature that satisfies an information need within large collection. In addition, actually, several problems have become closely related to information retrieval due the large amount of information scattered in many repositories. Many systems are relying on MAS due their intelligent and distributed nature. MAS can be considered as the more adapted approach to information retrieval problem. In the following, we discuss several systems based on MAS technology for information retrieval.

Amalthaes [48] is based on MAS for personalized, filtering and monitoring information. It consists of assisting user in finding interesting information by using to kind of agents: (i) filtering agent that model and monitor the interest of user, (ii) discovery agent that model the information sources.

Letizia [49] is relying on behavior based interface agent that assists a user browsing the World Wide Web. It aims to use the past behavior of users to anticipate a rough approximation of the user's interests.

WebWatcher [50] is an information seeking agent for WWW that assists user locate desired information by finding hyperlinks which are likely to lead to the targetinformation. It uses the stored previous searches combined to machine learning methods to return the appropriate hyperlink to user's goal.

SoftBot [51] is a system that accepts high-level user goals and dynamically synthesizes the appropriate sequence of Internet commands to satisfy those goals.

Retsina [52] is an implemented MAS infrastructure that have been applied in many domains. warren is an example of Retsina application that based on three types of agents (i) interface agent (ii) task agent that assist the user and (iii) information agents that re used to gather relevant information;

IR agents [53] is a multi-agents model for information retrieval on WWW. An IR agent rely on three types of agent: (i) Managing agents for extracting the semantics of information and managing the details of co-ordinate agents, (ii) Interface agents for interacting between the system and users and (iii) Search agents for discovering the information on WWW.

In CEMAS [54] (Concept ExchangingMulti-Agent System) the main focus is to provide specialized agent for each main task. It defines four main agents: (i) Broker agent is a central depository for resource and service knowledge in the multi-agent system, (ii) client agent as an interface to the system, (iii) search agent for new information and (iv) Server agent manages all the knowledge about concepts and links.

NetSa [55] (Networked Software Agents) is a MAS for Internet retrieval. It relies on five types of agents: broker agent, user agent, execution agent and resource agent. These agents are grouped in three main units: (i) communication Unit take care of communication between users and the system, (ii) processing information unit receive request from communication unit and decompose it on several simple requests forwarded to extraction unit, (iii) extraction unit is an interface between system and repositories.

XMAS [67] is a generic Architecture aimed at retrieving, filtering and reorganizing information according to user interests. It is stratified in four levels: (1) information level, (2) filter level, (3) Task level and (4) interface level.

IV. PROPOSED MODEL

The proposed system is based on multimodal approaches. It combines several methods throughout the information lifecycle since crawling in the web, until ranking the result delivery. Indeed, we believe that video retrieval information is a collaborative process where the overall system performance is closely linked to all information lifecycle phases. As known, the ideal performance is 1. For the global performance (P), it is equal to the multiplication of the local performances (Pi).

$$P = \prod n\, Pi$$

In the focus to ameliorate our system performance, we divided the system into independent parts easy to develop and maintain thereafter. We also involve the end user in the continual improving performance, and thus the quality and relevance of the results returned by using an active feedback. In the following, we give a detailed description of the system in layers at first, and then, we provide the list of used agents in the system.

*A. system layers*

In order to answer users' expectations in the best conditions with an acceptable quality, our model is based on multi-agents system which provides required services needed to increase the system efficiency. The overall system architecture consists of an agents' community grouped into four layers that communicate with each other displayed as follow:

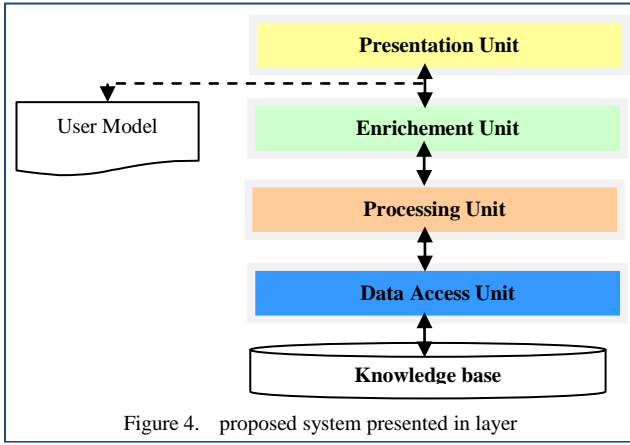

Figure 4. proposed system presented in layer

*1) Presentation layer*

The presentation layer is considered as the user-side system. In fact, this layer is located in the upper level that's, on the one hand allows users to exchange data with a personalized manner; on the other hand it presents different entries to express the users' needs.

*2) Data access layer*

The goal of the data access layer is to allow an easy way to access data. In fact, the data access layer provides a set of tools which simplified access to the data stored in the knowledge base. It can be accessed through the upper layers with different ways either directly by using the raw returned results or indirectly via an intermediate layer by providing processed data. In addition, the present layer is based on transparency transactions which hide all the complexity that lies behind this fundamental task.

*3) Processing layer*

Processing layer consists of take refining request from enrichment layer and maps it with a specific knowledge domain designed by user before starting retrieval task. The mapping task use similarity between terms in refined request and specific domain ontology concept. As seen in figure 2, the adopted approach to perform this task is to match each term in request with ontology concept via WordNet [56].

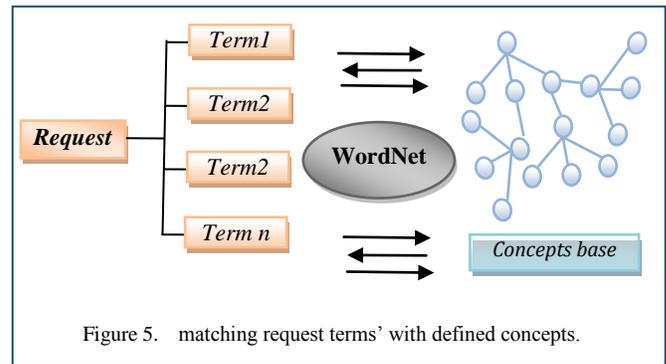

Figure 5. matching request terms' with defined concepts.

The main role of this layer is to add user preferences' with the raw user request. This layer will help the system to discriminate noisy information and to increase semantics in user request. It consist of inject meaningful terms related to the specific domain for a better accuracy.

*B. System Agents*

*1) Avatar agent:*

The avatar agent it can be considered as the heart of the system. In fact the avatar agent takes a leader place on the agent society. It is the representative of the user within the system. Each user has a unique avatar agent which has the necessary rights to access the whole personal information. He is able to manage the eventual changes of states of this information. Any interaction between the user and the system imperatively passes by the avatar agent. Moreover, it is brought to collect the relevant information produced by the user as the satisfaction degree of the returned answers and those regarding the new trends/preferences of the user while keeping the links preference/context. By context we mean the triplet location, time and device.

When he is created, an avatar agent has no experience or preference able to help it to achieve its goal. In order to minimize the time for gathering information and learning preferences, by default, an avatar agent will be affected directly at one or several communities. Their preferences will constitute the basic preferences of our agent. The most intuitive community will be a geographical community (country) and/or linguistics one.

*2) Facet agent*

When searching for information in the wide domain video was a rough and delicate task, it was wise to proceed to a classification into several domains for a better surrounding of this large space from the perspective of "divide to reign". We talk about narrow domain described, Smeulders et al[17], as having "a limited variability in all relevant aspects of its appearance. Contrary to a narrow, the wide domain "has an unlimited and unpredictable variability in its appearance even for the same semantic meaning". Smeulders et al[17]. Therefore, to meet this need, the facet agent is essential. In fact, to have a specific vision for a given domain (news, sports, art…) a facet agent aims to deal with characteristics (trends/preferences) of each one (figure) using the appropriate user domain. To this end, each user possesses its own strategy or kind research manner for each domain. Therefore, in order to answer to this task, our system provides a set of predefined "search strategies". The strategy attribution process is managed by the "Strategist Agent" which will be detailed later. Thus, the facet agent assists its "master" when he began a search session. It gives suggestions to the user from the relevant information gathered before the last sessions. We distinguish two kinds of suggestions: (i) Search from the previous history, (ii)Predictive search;

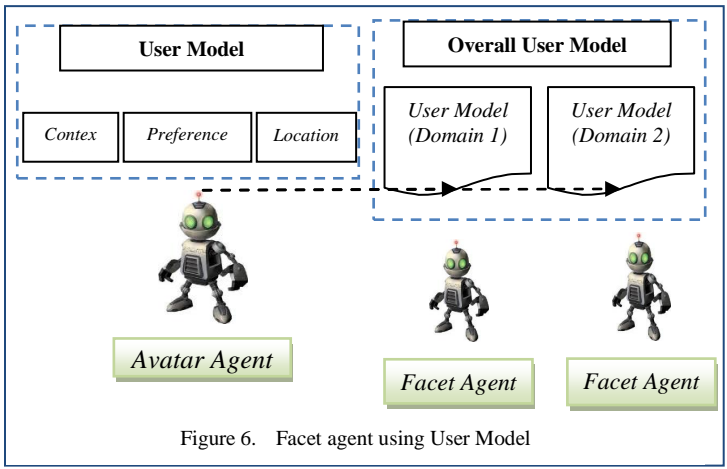

Figure 6. Facet agent using User Model

*3) Strategist Agent*

Since the system has a set of research strategies, an agent strategy aim to choose the most appropriate strategy to a facet agent of the user in terms. The default strategy assigned to a new "facet agent" is identical to the most used Strategy by the facet agents related to users who have a common points. In our approach we focus on communities of belonging and their preferences.(Multiple feedback techniques integrated with a recommendation system to assist users in choosing the proper feedback technique;[Luan & al. 2008])

*4) Community Agent:*

By definition [oxford dictionary], a community is a group of people living in the same place or having a particular characteristic in common. In our case, we use the term of "community" to define a group of individuals who share points or common interests. Indeed, a point of interest can be a value, a need or intention. It can be established according to several criteria (geographic, linguistic, ethnic, religious ...).

A community agent is an agent who takes care of gathering, then aggregating information about a community to build a knowledge base of relevant information of their common interests. A community agent is considered as the entity which capitalize the experiences of every community member in order to share their common interests and to help one other to accomplish its ultimate purpose of finding the desired information with an easy way.

As seen before, each user is represented by an avatar agent who adheres to one or more communities. Every avatar agent may join several communities with a degree of membership on each of them. In addition to the user profile, the indicator of belonging will contribute greatly to establish the priority of each user interest.

*5) Agent crawler:*

Agent crawler has as mission to look for newest uploaded video in the web. It seeks a predefined list of repositories or web sites. Also, it's able to browse the web page using links within the pages. At the end, it returns a list of video links that will be extracted (figure3).

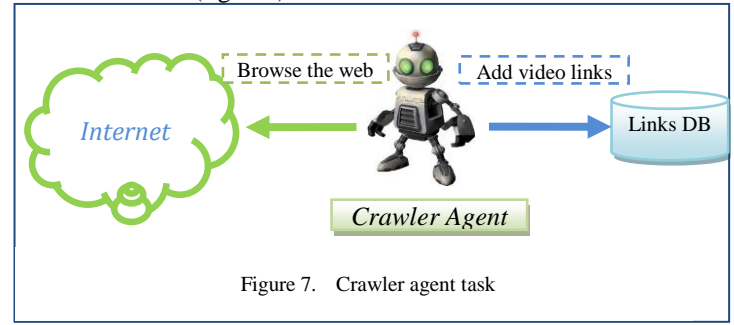

Figure 7. Crawler agent task

*6) Extractor agent*

Extractor agent extracts latest video links added to "Links DB". It downloads video related to each link. After segmenting video into shots and key frame, extractor agent extracts visual features (color, edge,…) and textual features (speech and captions) from each shot and frame caption. All metadata description will be formatted in MPEG7 (figure).

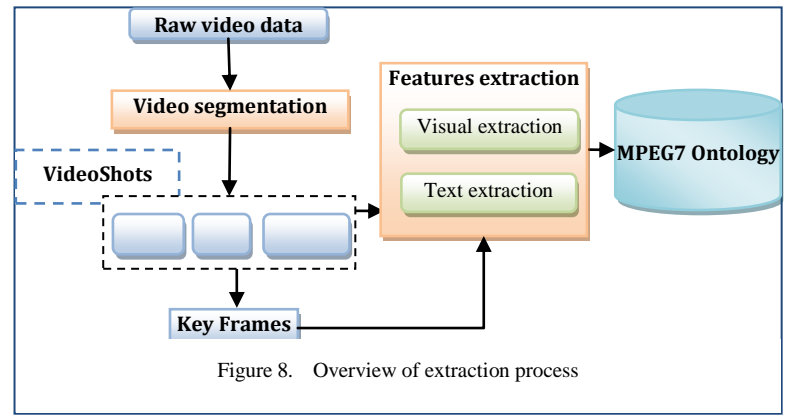

Figure 8. Overview of extraction process

*7) Classifier agent*

The classifier agent classify each new video or video shots with a predefined training concept using supervised learning SVM. It will match each new video to the closely concept for each domain (figure). The classification task is based on textual and visual information.It's an important phase in the lifecycle video information. It influences considerably each user results, thus, system performance.

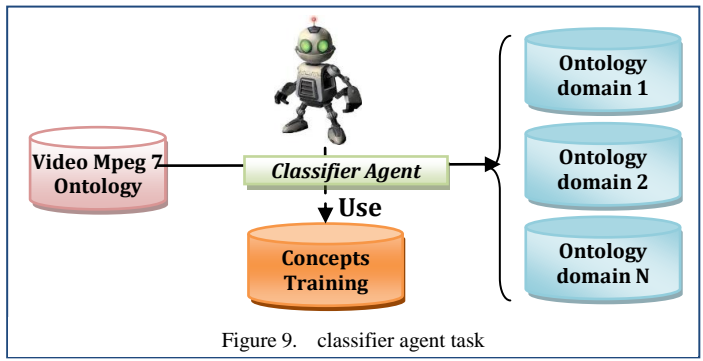

Figure 9. classifier agent task

Our knowledge base will be divided in two levels: the first level can be considered as a common knowledge base to all specific domains. It contains a multimedia ontology based on MPEG7 descriptors extracted from each stored video. The second level contains ontology specific to each domain; it will constitute the highly semantic level.

To improve the performance of the knowledge base, the system will conduct a filtering data by dividing it into three parts (figure 11): (i) Active base,(ii) usual base and (iii) depreciated base. The active base contains the most requested documents according to the preferences of all users. It will allow discriminating the noise generated by no-relevant documents to all user queries in the future as was in the past. Usual base is the knowledge base which contains documents which have an acceptable ranking. It's the second level of the overall knowledge when deprecated base is the third level which contains no relevant documents for all users. When the system start for the first time all stored documents are eventually relevant for users. After each result, user ranks the returned documents.

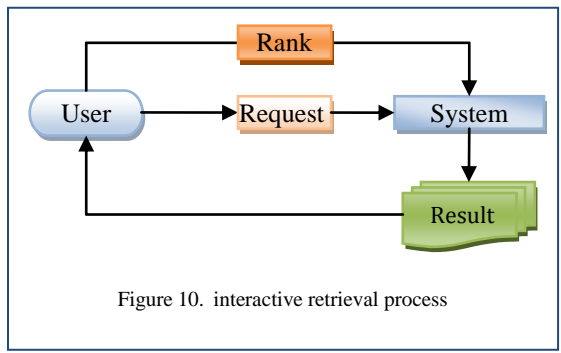

Figure 10. interactive retrieval process

*8) Organizer agent*

In order to organize the overall knowledge base, organizer agent based onant colony algorithm, will change the position of document between the three levels of knowledge base. In this case, each document constitutes an eventual desired food result.Depending on time and number of user requests, each document not desired by user will be considered as too far from the ant nest.Thus, reorganization of the knowledge base will be done adopting this logic.

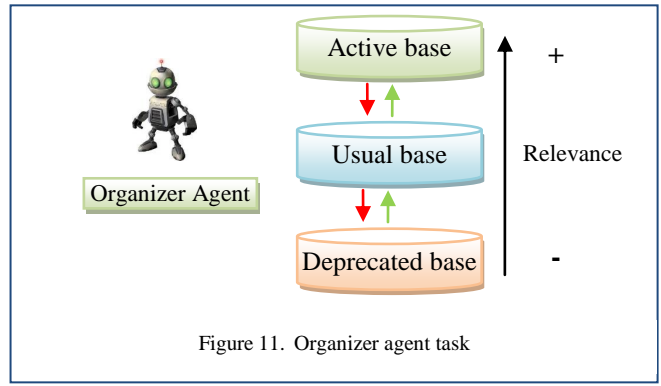

Figure 11. Organizer agent task

V. CONCLUSIONS AND FITURE WORKS

In this paper we have proposed a newest model on extracting, indexing and retrieving video information from the web based on multimodal approaches in semantic manner. It takes into account all steps in the lifecycle of retrieving information since extraction until result delivery; as future we plan to provide a prototype as implementation of this solution and we intend to measure it performance.

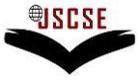

Citation:

Yasser El Madani El Alami, El Habib Nfaoui, Omar El Beqqali"Multi-agents Architecture for Semantic
Retrieving Video in Distributed Environment", International Journal of Soft Computing and Software Engineering [JSCSE],
Vol. 3, No. 3, pp. 430-440, 2013